\documentclass[aps,prd,twocolumn,superscriptaddress]{revtex4}
\usepackage{epsfig,epsf}
\usepackage{amsmath}
\usepackage{amsthm}
\usepackage{amsfonts}
\usepackage{amssymb}
\usepackage{dsfont}
\usepackage{multirow}
\usepackage{appendix}
\usepackage{slashed}
\usepackage[active]{srcltx}
\usepackage{psfrag}

\begin{document}

\title{Strange partners of the doubly charmed tetraquark $T^{+}_{cc}$}
\date{\today}
\author{S.~S.~Agaev}
\affiliation{Institute for Physical Problems, Baku State University, Az--1148 Baku,
Azerbaijan}
\author{K.~Azizi}
\thanks{Corresponding Author}
\affiliation{Department of Physics, University of Tehran, North Karegar Avenue, Tehran
14395-547, Iran}
\affiliation{Department of Physics, Do\v{g}u\c{s} University, Dudullu-\"{U}mraniye, 34775
Istanbul, Turkey}
\author{H.~Sundu}
\affiliation{Department of Physics, Kocaeli University, 41380 Izmit, Turkey}
\affiliation{Department of Physics Engineering, Istanbul Medeniyet University, 34700
Istanbul, Turkey}

\begin{abstract}
The spectroscopic parameters and widths of the axial-vector $T_{cc:s}^{%
\mathrm{AV}}$ and scalar $T_{cc:s}^{\mathrm{S}}$, $\widetilde{T}_{cc:s}^{%
\mathrm{S}}$ strange partners of the doubly charmed exotic meson $T_{cc}^{+}$
with the content $cc\overline{u}\overline{s}$, are calculated in the
framework of the QCD sum rule method. We model $T_{cc:s}^{\mathrm{AV}}$ as
the diquark-antidiquark state composed of axial-vector and scalar
components, whereas scalar particles $T_{cc:s}^{\mathrm{S}}$ and $\widetilde{%
T}_{cc:s}^{\mathrm{S}}$ are built of axial-vector and scalar diquarks,
respectively. The masses and current couplings of these tetraquarks are
calculated in the context of the two-point sum rule approach by taking into
account the quark, gluon and mixed condensates up to dimension $10$. The
full width of the state $T_{cc:s}^{\mathrm{AV}}$ is found from analysis of
the processes $T_{cc:s}^{\mathrm{AV}} \to D^{0}D_{s}^{\ast +}$ and $%
T_{cc:s}^{\mathrm{AV}} \to D^{\ast}(2007)^{0}D_{s}^{+}$. Decays to $%
D^{0}D_{s}^{+}$, $D^{\ast}(2007)^{0}D_{s}^{\ast+}$ and $D^{0}D_{s}^{+}$
mesons are utilized in the case of the scalar tetraquarks $T_{cc:s}^{\mathrm{%
S}}$ and $\widetilde{T}_{cc:s}^{\mathrm{S}}$, respectively. The partial
widths of the aforementioned decays are determined via the strong couplings $%
g_1$, $g_2 $, $G_1$, $G_2$ and $\widetilde{G}$, which describe the strong
interactions of the particles at the relevant tetraquark-meson-meson
vertices. These couplings are computed using the QCD three-point sum rule
method, most appropriate for the strong decays under study. The predictions $%
m=(3995 \pm 143)~\mathrm{MeV} $ and $\Gamma_{\mathrm{AV}}=(72 \pm 13)~%
\mathrm{MeV} $, as well as $m_{\mathrm{S}}=(4128 \pm 142)~\mathrm{MeV}$, $%
\Gamma_{\mathrm{S}}=(213 \pm 47)~\mathrm{MeV}$ and $\widetilde{m}_{\mathrm{S}%
}=(4035 \pm 145)~\mathrm{MeV} $ and $\widetilde{\Gamma}_{\mathrm{S}}=(123
\pm 32)~\mathrm{MeV} $ obtained for the masses and widths of these
tetraquarks in the present work can be useful in future experimental
investigations of the doubly charmed four-quark resonances.
\end{abstract}

\maketitle


\section{Introduction}

\label{sec:Int}
Recently, the LHCb Collaboration discovered the doubly charmed axial-vector
four-quark meson $T_{cc}^{+}$ with content $cc\overline{u}\overline{d}$ \cite%
{Aaij:2021vvq,LHCb:2021auc}. It was observed as a narrow peak in $%
D^{0}D^{0}\pi ^{+}$ invariant mass distribution, and is the first doubly
charmed state seen in the experiment. The state $T_{cc}^{+}$ has interesting
features: its mass $m_{\exp }=3875.1~\mathrm{MeV}+\delta m_{\exp }$, where $%
\delta m_{\exp }=-273\pm 61\pm 5_{-14}^{+11}~\mathrm{KeV}$, is less than
two-meson $D^{0}D^{\ast }(2010)^{+}$ threshold , but is very close to it. It
is also the longest living exotic meson\ observed till now, because it has a
very small full width $\Gamma =410\pm 165\pm 43_{-38}^{+18}~\mathrm{KeV}$.

It is worth noting that the four-quark states $QQ\overline{q}\overline{q}%
^{\prime }$ containing two heavy quarks $Q$ were already in the center of
theoretical investigations \cite%
{Esposito:2013fma,Navarra:2007yw,Karliner:2017qjm,Eichten:2017ffp}. The
reason is that some of these tetraquarks may be stable against the strong
and electromagnetic decays, and therefore can transform to conventional
mesons only through weak interaction. Since widths of such particles would
be very small, they may be discovered in various exclusive and inclusive
processes relatively easily. In the case of tetraquarks $bb\overline{q}%
\overline{q}^{\prime }$ with different spin-parities and light antidiquark
contents $\overline{q}\overline{q}^{\prime }$, this assumption seems is
confirmed by numerous studies performed in the context of various models of
high energy physics. Thus, stability of the axial-vector tetraquark $%
T_{bb}^{-}=bb\overline{u}\overline{d}$ was proved in Refs.\ \cite%
{Navarra:2007yw,Karliner:2017qjm,Eichten:2017ffp,Agaev:2018khe}. In Ref.\
\cite{Agaev:2018khe} it was demonstrated that the mass of such particle, $%
m=(10035\pm 260)~\mathrm{MeV}$, is below the $B^{-}\overline{B}^{\ast 0}$
and $B^{-}\overline{B}^{0}\gamma $ thresholds. This means that the
double-beauty tetraquark $T_{bb}^{-}$ is stable against the strong and
radiative decays and dissociates to ordinary mesons only weakly. The full
width and mean lifetime of $T_{bb}^{-}$ were evaluated in Refs.\ \cite%
{Agaev:2018khe,Agaev:2020mqq} using various semileptonic and nonleptonic
decay channels of this particle. Results obtained for these parameters, $%
\Gamma _{\mathrm{full}}=(7.72\pm 1.23)\times 10^{-8}~\mathrm{MeV}$ and $\tau
=8.53_{-1.18}^{+1.57}~\mathrm{fs,}$ are useful for experimental
investigations of the tetraquark $T_{bb}^{-}$. Some of other $bb\overline{q}%
\overline{q}^{\prime }$ states were also identified as stable particles, and
their full widths were calculated via allowed weak transformations \cite%
{Xing:2018bqt,Agaev:2020zag,Agaev:2020dba,Agaev:2019lwh}.

The doubly charmed tetraquarks $cc\overline{q}\overline{q}^{\prime }$, as
interesting objects, were undergone of intensive studies as well \cite%
{Navarra:2007yw,Karliner:2017qjm,Eichten:2017ffp,Du:2012wp,Wang:2017uld,Wang:2017dtg,Braaten:2020nwp,Cheng:2020wxa, Meng:2020knc,Junnarkar:2018twb,Guo:2021yws,Padmanath:2022cvl}%
. Thus, the axial-vector tetraquark $cc\overline{u}\overline{d}$ was
explored in Ref.\ \cite{Navarra:2007yw} in the framework of QCD sum rule
method. Properties of the four-quark mesons $cc\overline{q}\overline{q}%
^{\prime }$ with spin-parities $J^{\mathrm{P}}=0^{-},~0^{+},~1^{-}$ and $%
1^{+}$ were analyzed in Ref.\ \cite{Du:2012wp}. New investigations of these
tetraquarks proved once more the unstable nature of the axial-vector
tetraquark $cc\overline{u}\overline{d}$ \cite%
{Karliner:2017qjm,Eichten:2017ffp,Wang:2017uld,Wang:2017dtg,Braaten:2020nwp,Cheng:2020wxa}%
. In contrast to these conclusions, studies carried out in the context of
the constituent quark model and lattice simulations demonstrated existence
of the stable axial-vector tetraquark $cc\overline{u}\overline{d}$ lying $%
\approx 23~\mathrm{MeV}$ below the two-meson threshold\ \cite%
{Meng:2020knc,Junnarkar:2018twb}.

Detailed investigations of doubly charmed tetraquarks were carried out also
in our articles \cite{Agaev:2019qqn,Agaev:2018vag}. In the first of these
publications, we calculated various parameters of pseudoscalar and scalar
states $cc\overline{u}\overline{d}$. Our analyses demonstrated that these
four-quark systems are unstable particles and break up to conventional
mesons trough strong interactions. Full widths of these particles were
evaluated using their kinematically allowed decays to $D^{+}D^{\ast
}(2007)^{0}$, $D^{0}D^{\ast }(2010)^{+}$ and $D^{0}D^{+}$ mesons,
respectively. Interesting hypothetical pseudoscalar tetraquarks $cc\overline{%
s}\overline{s}$ and $cc\overline{d}\overline{s}$ carrying two units of
electric charge were object of the second paper.

Experimental observation of the resonance $T_{cc}^{+}$ triggered new
analyses aimed to explain the measured parameters of this particle \cite%
{Agaev:2021vur,Feijoo:2021ppq,Yan:2021wdl,Fleming:2021wmk,Agaev:2022ast}. In
our papers \cite{Agaev:2021vur,Agaev:2022ast}, we addressed the relevant
problems in the framework of QCD sum rule method. In Ref.\ \cite%
{Agaev:2021vur}, the doubly charmed axial-vector resonance $T_{cc}^{+}$ was
treated as the diquark-antidiquark state $cc\overline{u}\overline{d}$. The
mass and coupling of this tetraquark were calculated by means of the QCD
two-point sum rule approach.

To estimate the full width of $T_{cc}^{+}$, we specified its decay channels
and used QCD three-point sum rule method to find the strong couplings at
relevant vertices. Because $T_{cc}^{+}$ is a very narrow state, its possible
decay modes were objects in numerous analyses. Indeed, $T_{cc}^{+}$ was
discovered in the $D^{0}D^{0}\pi ^{+}$ mass distribution. One of popular
ways to explain such final state is the chain of transformations $%
T_{cc}^{+}\rightarrow D^{0}D^{\ast +}\rightarrow D^{0}D^{0}\pi ^{+}$. But
the mass of $T_{cc}^{+}$ is below the $D^{0}D^{\ast +}$ threshold, therefore
the process $T_{cc}^{+}\rightarrow D^{0}D^{\ast +}$ is kinematically
forbidden, and may proceed only virtually. Alternatively, production of $%
D^{0}D^{0}\pi ^{+}$ may run through a process $T_{cc}^{+}\rightarrow T_{cc;%
\overline{u}\overline{u}}^{0}\pi ^{+}$ where $T_{cc;\overline{u}\overline{u}%
}^{0}$ is a scalar tetraquark. Then three-meson final state may appear due
to the decay $T_{cc;\overline{u}\overline{u}}^{0}\rightarrow D^{0}D^{0}$. To
calculate the full width of $T_{cc}^{+}$, we explored also another mode $%
T_{cc}^{+}\rightarrow \widetilde{T}_{cc;\overline{u}\overline{d}}^{+}\pi
^{0}\rightarrow D^{0}D^{+}\pi ^{0}$ with $\widetilde{T}_{cc;\overline{u}%
\overline{d}}^{+}$ being a scalar tetraquark. Results found for the mass and
width of the diquark-antidiquark state $T_{cc}^{+}$ are in nice agreement
with the LHCb data.

The hadronic molecule model $D^{0}D^{\ast +}$ for the resonance $T_{cc}^{+}$
was examined in Ref.\ \cite{Agaev:2022ast}. It turned out that the mass and
width of this system exceed experimental data of the LHCb Collaboration. Our
studies showed that a preferable assignment for the LHCb resonance $%
T_{cc}^{+}$ is the diquark-antidiquark model, but because parameters of the
molecule $D^{0}D^{\ast +}$ suffer from theoretical uncertainties, we did not
exclude the molecule picture for $T_{cc}^{+}$.

The four-quark states containing strange $s$-quark alongside the heavy
diquark $cc$ form another interesting class of the doubly charmed particles.
These exotic mesons may have diquark-antidiquark or molecular structures. In
the context of various methods, masses of the doubly charmed strange
tetraquarks with different spin-parities and structures were considered in
Refs.\ \cite%
{Karliner:2021wju,Chen:2022ros,Praszalowicz:2022sqx,Xin:2021wcr,Ren:2021dsi}.

In the present article, we are going to explore features of axial-vector $%
T_{cc:s}^{\mathrm{AV}}$ and scalar $T_{cc:s}^{\mathrm{S}}$, $\widetilde{T}%
_{cc:s}^{\mathrm{S}}$ diquark-antidiquark systems $cc\overline{u}\overline{s}
$ (in what follows, $T_{\mathrm{AV}}$, $T_{\mathrm{S}}$ and $\widetilde{T}_{%
\mathrm{S}}$, respectively) by computing their masses, current couplings and
widths. We model the tetraquark $T_{\mathrm{AV}}$ as a four-quark meson
composed of a heavy axial-vector diquark and light scalar antidiquark. The
scalar tetraquarks $T_{\mathrm{S}}$ and $\widetilde{T}_{\mathrm{S}}$ are
considered as particles built of the axial-vector and scalar constituents,
respectively. To calculate the masses and couplings of these structures, we
use QCD two-point sum rules by taking into account various vacuum
condensates up to dimension $10$. The width of $T_{\mathrm{AV}}$ is
evaluated by considering its $S$-wave decays $T_{\mathrm{AV}}\rightarrow
D^{0}D_{s}^{\ast +}$ and $T_{\mathrm{AV}}\rightarrow D^{\ast
}(2007)^{0}D_{s}^{+}$. The widths of scalar tetraquarks $T_{\mathrm{S}}$ and
$\widetilde{T}_{\mathrm{S}}$ are estimated using their decays to $%
D^{0}D_{s}^{+}$, $D^{\ast }(2007)^{0}D_{s}^{\ast +}$ and $D^{0}D_{s}^{+}$
mesons, respectively. Partial widths of these processes are determined by
the strong couplings $g_{1}$, $g_{2}$, $G_{1}$, $G_{2}$ and $\widetilde{G}$
at tetraquark-meson-meson vertices $T_{\mathrm{AV}}D^{0}D_{s}^{\ast +}$, $T_{%
\mathrm{AV}}D^{\ast }(2007)^{0}D_{s}^{+}$, $T_{\mathrm{S}}D^{0}D_{s}^{+}$, $%
T_{\mathrm{S}}D^{\ast }(2007)^{0}D_{s}^{\ast +}$and $\widetilde{T}_{\mathrm{S%
}}D^{0}D_{s}^{+}$, respectively.

This article is structured in the following manner: In Sec.\ \ref{sec:Masses}%
, we calculate the mass and current coupling of the tetraquarks $T_{\mathrm{%
AV}}$, $T_{\mathrm{S}}$ and $\widetilde{T}_{\mathrm{S}}$ in the framework of
QCD two-point sum rule method. The strong couplings $g_{1}$, $g_{2}$ and
partial widths of the decays $T_{\mathrm{AV}}\rightarrow D^{0}D_{s}^{\ast +}$
and $T_{\mathrm{AV}}\rightarrow D^{\ast }(2007)^{0}D_{s}^{+}$, as well as
the full width of $T_{\mathrm{AV}}$ are found in Sec.\ \ref{sec:Width1}.
Section \ref{sec:Width2} is devoted to computation of the strong couplings$\
G_{1}$, $G_{2}$ and $\widetilde{G}$ and widths of the tetraquarks $T_{%
\mathrm{S}}$ and $\widetilde{T}_{\mathrm{S}}$. Section\ \ref{sec:Conclusion}
is reserved for discussion ans concluding notes.


\section{Spectroscopic parameters of the tetraquarks $T_{\mathrm{AV}}$, $T_{%
\mathrm{S}}$, and $\widetilde{T}_{\mathrm{S}}$}

\label{sec:Masses}

In this section, we calculate the masses and current couplings of the
tetraquarks $T_{\mathrm{AV}}$, $T_{\mathrm{S}}$, and $\widetilde{T}_{\mathrm{%
S}}$ using the two-point sum rule method \cite{Shifman:1978bx,Shifman:1978by}%
. The key component in the sum rule analysis is an interpolating current for
a hadron under consideration. In the diquark-antidiquark model, the
four-quark mesons $T_{\mathrm{AV}}$, $T_{\mathrm{S}}$, and $\widetilde{T}_{%
\mathrm{S}}$ are built of the heavy diquark $cc$ and light antidiquark $%
\overline{u}\overline{s}$. In the case of the axial-vector state $T_{\mathrm{%
AV}}$, we suggest that it is composed of the axial-vector diquark $%
c^{T}C\gamma _{\mu }c$ and scalar antidiquark $\overline{u}\gamma _{5}C%
\overline{s}^{T}$ with $C$ being the charge conjugation matrix. We take into
account that the axial-vector diquark $c_{a}^{T}C\gamma _{\mu }c_{b}$, where
$a$ and $b$ are color indices, has a symmetric flavor but an antisymmetric
color organization, and its flavor-color structure is $(\mathbf{6}_{f},%
\overline{\mathbf{3}}_{c})$ \cite{Du:2012wp}. Then, the color-singlet
current $J_{\mu }(x)$ for $T_{\mathrm{AV}}$ can be constructed using the
color-triplet light antidiquark field $\overline{u}_{a}\gamma _{5}C\overline{%
s}_{b}^{T}-\overline{u}_{b}\gamma _{5}C\overline{s}_{a}^{T}$. Because both
components of this field lead to identical terms in $J_{\mu }(x)$, it is
enough to preserve in calculations one of them. As a result, for the
current\ $J_{\mu }(x)$,\ we get the following expression
\begin{equation}
J_{\mu }(x)=\left[ c_{a}^{T}(x)C\gamma _{\mu }c_{b}(x)\right] \left[
\overline{u}_{a}(x)\gamma _{5}C\overline{s}_{b}^{T}(x)\right],
\label{eq:CurrVector}
\end{equation}%
which belongs to the $[\overline{\mathbf{3}}_{c}]_{cc}\otimes \lbrack
\mathbf{3}_{c}]_{\overline{u}\overline{s}}$ representation of the color
group $SU_{c}(3)$.

In the case of the scalar exotic meson $cc\overline{u}\overline{s}$, we
consider two structures $T_{\mathrm{S}}$ and $\widetilde{T}_{\mathrm{S}}$
which can describe this tetraquark. We assume that $T_{\mathrm{S}}$ is a
scalar tetraquark built of axial-vector components. The interpolating
current for such state is given by the expression
\begin{equation}
J(x)=[c_{a}^{T}(x)C\gamma _{\mu }c_{b}(x)][\overline{u}_{a}(x)\gamma ^{\mu }C%
\overline{s}_{b}^{T}(x)],  \label{eq:Curr1}
\end{equation}%
which is from the antitriplet-triplet representation of the color group $%
SU_{c}(3)$.

We model $\widetilde{T}_{\mathrm{S}}$ by supposing that it is made of the
scalar diquark $c_{a}^{T}C\gamma _{5}c_{b}$ and antidiquark $\overline{u}%
_{a}\gamma _{5}C\overline{s}_{b}^{T}$. The heavy diquark has symmetric
flavor and color structure $(\mathbf{6}_{f},\mathbf{6}_{c})$, i.e., has a
color-flavor sextet organization \cite{Du:2012wp}. Then the light
antidiquark should be also symmetric in the color indices $\overline{u}%
_{a}\gamma _{5}C\overline{s}_{b}^{T}+\overline{u}_{b}\gamma _{5}C\overline{s}%
_{a}^{T}$. The interpolating current $\widetilde{J}(x)$ of such tetraquark
contains relevant diquark and antidiquark fields and has $[\mathbf{6}%
_{c}]_{cc}\otimes \lbrack \overline{\mathbf{6}}_{c}]_{\overline{u}\overline{s%
}}$ color-structure. In investigations, we employ
\begin{equation}
\widetilde{J}(x)=[c_{a}^{T}(x)C\gamma _{5}c_{b}(x)][\overline{u}%
_{a}(x)\gamma _{5}C\overline{s}_{b}^{T}(x)],  \label{eq:CurrScalar}
\end{equation}%
because components of the antidiquark field generate in the current $%
\widetilde{J}$ two equal terms.


\subsection{The mass and coupling of \ $T_{\mathrm{AV}}$}

\label{subsec:SS1}
To extract sum rules for the spectroscopic parameters of the tetraquark $T_{%
\mathrm{AV}}$, we begin from analysis of the correlation function $\Pi _{\mu
\nu }(p)$,
\begin{equation}
\Pi _{\mu \nu }(p)=i\int d^{4}xe^{ipx}\langle 0|\mathcal{T}\{J_{\mu
}(x)J_{\nu }^{\dag }(0)\}|0\rangle .  \label{eq:CF1}
\end{equation}%
First, $\Pi _{\mu \nu }(p)$ should be expressed using the mass $m$ and
coupling $f$ of the tetraquark $T_{\mathrm{AV}}$: This expression is the
hadronic representation of the $\Pi _{\mu \nu }(p)$ and forms the physical
side of the desired sum rules. To this end, we saturate the correlation
function $\Pi _{\mu \nu }(p)$ with a complete set of states with quantum
numbers $J^{\mathrm{P}}=1^{+}$ and perform the integration over $x$ in Eq.\ (%
\ref{eq:CF1}),
\begin{equation}
\Pi _{\mu \nu }^{\mathrm{Phys}}(p)=\frac{\langle 0|J_{\mu }|T_{\mathrm{AV}%
}(p,\epsilon )\rangle \langle T_{\mathrm{AV}}(p,\epsilon )|J_{\nu }^{\dagger
}|0\rangle }{m^{2}-p^{2}}+\cdots .  \label{eq:PhysSide}
\end{equation}%
We write down the contribution arising from the ground-state particle $T_{%
\mathrm{AV}}$ explicitly, and denote ones due to higher resonances and
continuum states by the dots.

The function $\Pi _{\mu \nu }^{\mathrm{Phys}}(p)$ can be simplified using
the matrix element,
\begin{equation}
\langle 0|J_{\mu }|T_{\mathrm{AV}}(p,\epsilon )\rangle =fm\epsilon _{\mu },
\label{eq:MElem1}
\end{equation}%
where $\epsilon _{\mu }$ is the polarization vector of $T_{\mathrm{AV}}$. It
is not difficult to demonstrate that in terms of $m$ and $f$ the function $%
\Pi _{\mu \nu }^{\mathrm{Phys}}(p)$ takes the following form:
\begin{equation}
\Pi _{\mu \nu }^{\mathrm{Phys}}(p)=\frac{m^{2}f^{2}}{m^{2}-p^{2}}\left(
-g_{\mu \nu }+\frac{p_{\mu }p_{\nu }}{m^{2}}\right) +\cdots.
\label{eq:PhysSide1}
\end{equation}

The QCD side of the sum rules $\Pi _{\mu \nu }^{\mathrm{OPE}}(p)$ has to be
computed in the operator product expansion ($\mathrm{OPE}$) with some fixed
accuracy. The function $\Pi _{\mu \nu }^{\mathrm{OPE}}(p)$ can be obtained
by inserting the explicit form of the interpolating current $J_{\mu }(x)$
into Eq.\ (\ref{eq:CF1}), and contracting corresponding heavy and light
quark fields. After these operations, we get
\begin{eqnarray}
&&\Pi _{\mu \nu }^{\mathrm{OPE}}(p)=i\int d^{4}xe^{ipx}\mathrm{Tr}\left[
\gamma _{5}\widetilde{S}_{s}^{b^{\prime }b}(-x)\right.  \notag \\
&&\left. \times \gamma _{5}S_{u}^{a^{\prime }a}(-x)\right] \left\{ \mathrm{Tr%
}\left[ \gamma _{\nu }\widetilde{S}_{c}^{ba^{\prime }}(x)\gamma _{\mu
}S_{c}^{ab^{\prime }}(x)\right] \right.  \notag \\
&&\left. -\mathrm{Tr}\left[ \gamma _{\nu }\widetilde{S}_{c}^{aa^{\prime
}}(x)\gamma _{\mu }S_{c}^{bb^{\prime }}(x)\right] \right\} .
\label{eq:QCDside}
\end{eqnarray}%
In Eq.\ (\ref{eq:QCDside}), $S_{c}^{ab}(x)$ and $S_{q}^{ab}(x)$ are the $c$
and $q(u,s)$-quark propagators: their analytic expressions were presented in
Ref.\ \cite{Agaev:2020zad}. We have also introduced the notation
\begin{equation}
\widetilde{S}_{c(q)}(x)=CS_{c(q)}^{T}(x)C.  \label{eq:Prop}
\end{equation}

The QCD sum rules are obtained by equating the invariant amplitudes which
correspond to the same Lorentz structures both in $\Pi _{\mu \nu }^{\mathrm{%
Phys}}(p)$ and $\Pi _{\mu \nu }^{\mathrm{OPE}}(p)$. The amplitudes
corresponding to the structures $\sim g_{\mu \nu }$ do not receive
contributions from the spin-$0$ particles, therefore, they are suitable for
our purposes. Having denoted these invariant amplitudes by $\Pi ^{\mathrm{%
Phys}}(p^{2})$ and $\Pi ^{\mathrm{OPE}}(p^{2})$, respectively, and equating
them, we find an expression which is convenient for further processing.
First, we encounter here problems connected with necessity to suppress the
contributions of the higher resonances and continuum states. For these
purposes, we make use of the Borel transformation in the sum rule equality.
Afterwards, employing the assumption about quark-hadron duality, it is
possible to subtract these effects from the obtained expression. After these
manipulations, the sum rule equality acquires dependence on the Borel $M^{2}$
and continuum threshold $s_{0}$ parameters.

It is evident that the Borel transformation of $\Pi ^{\mathrm{Phys}}(p^{2})$
is trivial. The Borel transformation and continuum subtraction convert the
invariant amplitude $\Pi ^{\mathrm{OPE}}(p^{2})$ to the form,
\begin{equation}
\Pi (M^{2},s_{0})=\int_{\mathcal{M}^{2}}^{s_{0}}ds\rho ^{\mathrm{OPE}%
}(s)e^{-s/M^{2}}+\Pi (M^{2}),  \label{eq:InvAmp}
\end{equation}%
where $\mathcal{M}=2m_{c}+m_{s}$. In numerical computations, we set $m_{u}=0$
and $m_{s}^{2}=0$, but take into account terms proportional to $m_{s}$. In
Eq.\ (\ref{eq:InvAmp}), $\rho ^{\mathrm{OPE}}(s)$ is the two-point spectral
density computed as the imaginary part of the correlation function. The
second component of the invariant amplitude $\Pi (M^{2})$ includes
nonperturbative contributions extracted directly from $\Pi _{\mu \nu }^{%
\mathrm{OPE}}(p)$. We calculate $\Pi (M^{2},s_{0})$ by taking into account
the nonperturbative terms up to dimension $10$. Explicit expression of $\Pi
(M^{2},s_{0})$ is rather cumbersome, therefore we do not provide it here.

The sum rules for $m$ and $f$ read
\begin{equation}
m^{2}=\frac{\Pi ^{\prime }(M^{2},s_{0})}{\Pi (M^{2},s_{0})},  \label{eq:Mass}
\end{equation}%
and
\begin{equation}
f^{2}=\frac{e^{m^{2}/M^{2}}}{m^{2}}\Pi (M^{2},s_{0}),  \label{eq:Coupling}
\end{equation}%
where $\Pi ^{\prime }(M^{2},s_{0})=d\Pi (M^{2},s_{0})/d(-1/M^{2})$.

The formulas in Eqs.\ (\ref{eq:Mass}) and (\ref{eq:Coupling}) depend on the
various quark, gluon and mixed condensates. They are universal quantities
extracted from numerous analyses:
\begin{eqnarray}
&&\langle \overline{q}q\rangle =-(0.24\pm 0.01)^{3}~\mathrm{GeV}^{3},\
\langle \overline{s}s\rangle =(0.8\pm 0.1)\langle \overline{q}q\rangle ,
\notag \\
&&\langle \overline{q}g_{s}\sigma Gq\rangle =m_{0}^{2}\langle \overline{q}%
q\rangle ,\ \langle \overline{s}g_{s}\sigma Gs\rangle =m_{0}^{2}\langle
\overline{s}s\rangle ,  \notag \\
&&m_{0}^{2}=(0.8\pm 0.2)~\mathrm{GeV}^{2}  \notag \\
&&\langle \frac{\alpha _{s}G^{2}}{\pi }\rangle =(0.012\pm 0.004)~\mathrm{GeV}%
^{4},  \notag \\
&&\langle g_{s}^{3}G^{3}\rangle =(0.57\pm 0.29)~\mathrm{GeV}^{6},  \notag \\
&&m_{s}=93_{-5}^{+11}~\mathrm{MeV},\ m_{c}=(1.27\pm 0.02)~\mathrm{GeV}.
\label{eq:Parameters}
\end{eqnarray}%
We also added the masses of $c$ and $s$ quarks to this list.

It is known that the Borel and continuum threshold parameters, $M^{2}$ and $%
s_{0}$, are auxiliary quantities of calculations, which have to satisfy some
constraints imposed on $\Pi (M^{2},s_{0})$ by the dominance of the pole
contribution ($\mathrm{PC}$) and convergence of the $\mathrm{OPE}$. We use
the following definition for $\mathrm{PC}$:
\begin{equation}
\mathrm{PC}=\frac{\Pi (M^{2},s_{0})}{\Pi (M^{2},\infty )}.  \label{eq:Pole}
\end{equation}%
The restriction on $\mathrm{PC}$ is necessary to fix the maximum value of
the Borel parameter. The low boundary of the window for the Borel parameter
is obtained from the convergence of the $\mathrm{OPE}$. To this end, we
employ the expression%
\begin{equation}
R(M^{2})=\frac{\Pi ^{\mathrm{DimN}}(M^{2},s_{0})}{\Pi (M^{2},s_{0})},
\label{eq:Convergence}
\end{equation}%
where $\Pi ^{\mathrm{DimN}}(M^{2},s_{0})$ is the contribution of the last
three terms in $\mathrm{OPE}$, i.e., $\mathrm{DimN=Dim(8+9+10)}$.

Apart from these constraints, the region for $M^{2}$ should lead to stable
predictions for extracted physical quantities. Our computations prove that
the regions for $M^{2}$ and $s_{0}$,
\begin{equation}
M^{2}\in \lbrack 3,4.5]~\mathrm{GeV}^{2},\ s_{0}\in \lbrack 20.5,21.5]~%
\mathrm{GeV}^{2},  \label{eq:Regions}
\end{equation}%
satisfy all restrictions mentioned above. In fact, in these regions, on
average in $s_{0}$, the pole contribution changes within the limits:
\begin{equation}
0.79\geq \mathrm{PC}\geq 0.49.  \label{eq:Polelimits}
\end{equation}%
At $M^{2}=3~\mathrm{GeV}^{2}$, the contributions to $\Pi (M^{2},s_{0})$ of
the last three terms in $\mathrm{OPE}$ is less than $0.01$.

\begin{widetext}

\begin{figure}[h!]
\begin{center}
\includegraphics[totalheight=6cm,width=8cm]{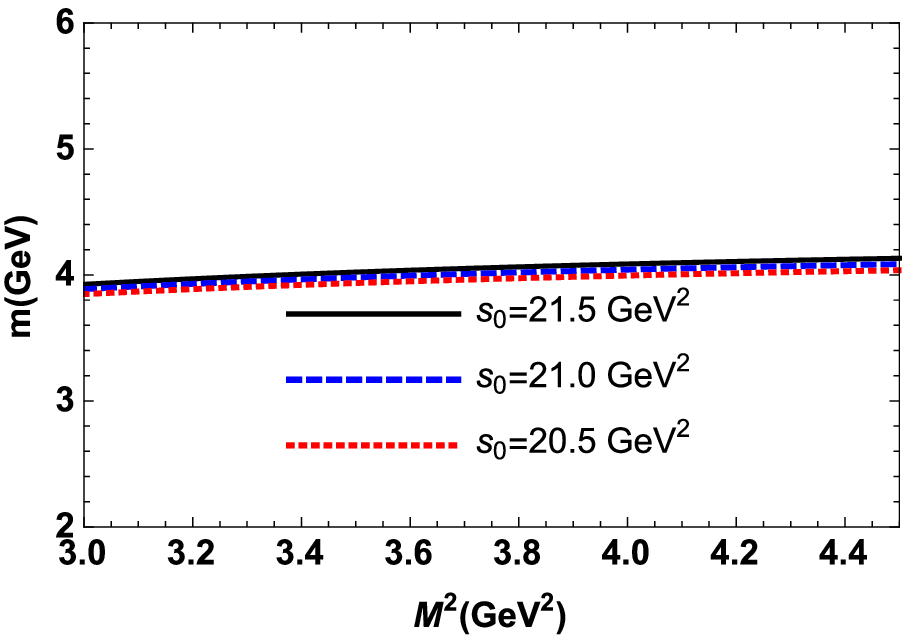}
\includegraphics[totalheight=6cm,width=8cm]{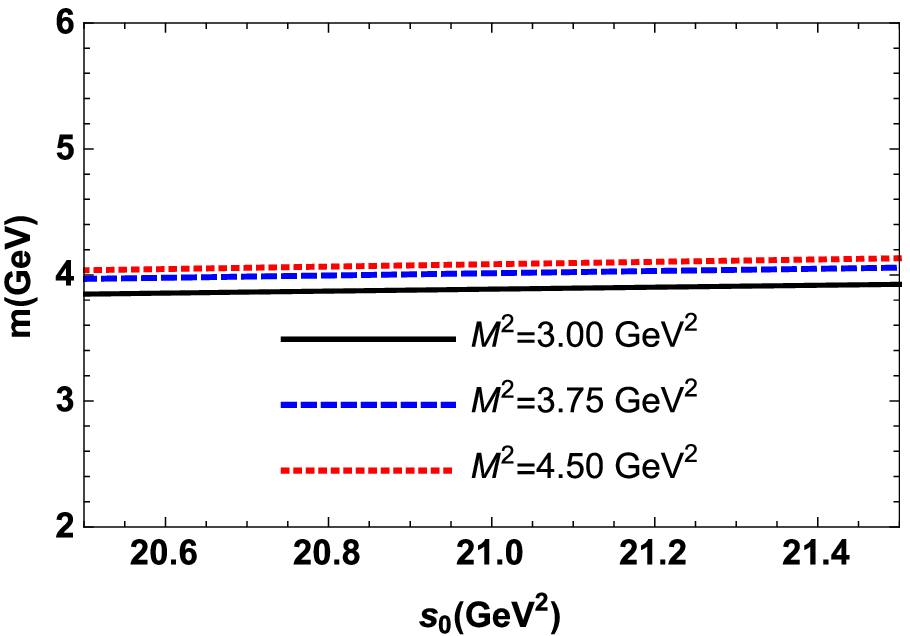}
\end{center}
\caption{Mass $m$ of the tetraquark $T_{\mathrm{AV}}$ as a function of the Borel parameter $M^{2}$ (left), and the continuum threshold parameter $s_0$ (right).}
\label{fig:Mass1}
\end{figure}

\end{widetext}

In Fig.\ \ref{fig:Mass1}, we plot the dependence of the mass $m$ of the
tetraquark $T_{\mathrm{AV}}$ on $M^{2}$ and $s_{0}$. It is clear, that the
window for $M^{2}$, where parameters of $T_{\mathrm{AV}}$ are extracted, can
be considered as a relatively stable plateau. Nevertheless, one sees a
residual dependence of $m$ on the Borel parameter $M^{2}$. This effect
allows one to find ambiguities of the sum rule calculations. Another source
of the theoretical uncertainties is the continuum threshold parameter $s_{0}$%
. The region for $s_{0}$ has to meet the constraints coming from the
dominance of $\mathrm{PC}$ and convergence of the $\mathrm{OPE}$. The
parameter $\sqrt{s_{0}}$ bears also information on the mass of the $T_{%
\mathrm{AV}}$ tetraquark's first radial excitation.

Our results for the spectral parameters of the tetraquark $T_{\mathrm{AV}}$
read
\begin{eqnarray}
m &=&(3995\pm 143)~\mathrm{MeV},  \notag \\
f &=&(2.4\pm 0.5)\times 10^{-3}~\mathrm{GeV}^{4}.  \label{eq:Result1}
\end{eqnarray}

The predictions for the mass $m$ and coupling $f$ are obtained as average of
results for these quantities in the working windows (\ref{eq:Regions}).
These values correspond to the sum rules' predictions approximately at
middle of these regions, i.e., to results at $M^{2}=3.5~\mathrm{GeV}^{2}$
and $s_{0}=21~\mathrm{GeV}^{2}$, where $\mathrm{PC}\approx 0.65$ guarantees
the ground-state nature of $T_{\mathrm{AV}}$. As it has been noted above, $%
\sqrt{s_{0}}-m\geq 480~\mathrm{MeV}$ allows us to estimate the mass of the
excited tetraquark as $m^{\ast }\approx (m+480)~\mathrm{MeV}$, which is
reasonable for double-heavy tetraquarks.


\subsection{Parameters of the tetraquarks $T_{\mathrm{S}}$ and $\widetilde{T}%
_{\mathrm{S}}$}

\label{subsec:SS2}

To find the mass $m_{\mathrm{S}}$ and current coupling $f_{\mathrm{S}}$ of
the scalar tetraquark $T_{\mathrm{S}}$, we start from the correlation
function $\Pi (p)$
\begin{equation}
\Pi (p)=i\int d^{4}xe^{ipx}\langle 0|\mathcal{T}\{J(x)J^{\dag
}(0)\}|0\rangle ,  \label{eq:CF1a}
\end{equation}%
where the current $J(x)$ is given by Eq.\ (\ref{eq:Curr1}). In terms of the
tetraquark's physical parameters, $\Pi (p)$ is determined by the expression,
\begin{equation}
\Pi ^{\mathrm{Phys}}(p)=\frac{\langle 0|J|T_{\mathrm{S}}(p)\rangle \langle
T_{\mathrm{S}}(p)|J^{\dagger }|0\rangle }{m_{\mathrm{S}}^{2}-p^{2}}+\cdots .
\label{eq:PhysSide2}
\end{equation}%
The function $\Pi ^{\mathrm{Phys}}(p)$ can be rewritten by employing the
matrix element,
\begin{equation}
\langle 0|J|T_{\mathrm{S}}(p)\rangle =m_{\mathrm{S}}f_{\mathrm{S}}.
\label{eq:MElem2}
\end{equation}%
Then, it is easy to see that in terms of the parameters $m_{\mathrm{S}}$ and
$f_{\mathrm{S}}$ the $\Pi ^{\mathrm{Phys}}(p)$ has the form,
\begin{equation}
\Pi ^{\mathrm{Phys}}(p)=\frac{m_{\mathrm{S}}^{2}f_{\mathrm{S}}^{2}}{m_{%
\mathrm{S}}^{2}-p^{2}}+\cdots .  \label{eq:PhysSide3}
\end{equation}

To obtain the QCD side of the sum rules, $\Pi ^{\mathrm{OPE}}(p)$, we insert
the interpolating current $J(x)$ into Eq.\ (\ref{eq:CF1a}), and contract the
relevant quark fields. After these manipulations, we get
\begin{eqnarray}
&&\Pi ^{\mathrm{OPE}}(p)=i\int d^{4}xe^{ipx}\mathrm{Tr}\left[ \gamma _{\mu }%
\widetilde{S}_{s}^{b^{\prime }b}(-x)\right.  \notag \\
&&\left. \times \gamma _{\nu }S_{u}^{a^{\prime }a}(-x)\right] \left\{
\mathrm{Tr}\left[ \gamma ^{\nu }\widetilde{S}_{c}^{aa^{\prime }}(x)\gamma
^{\mu }S_{c}^{bb^{\prime }}(x)\right] \right.  \notag \\
&&\left. -\mathrm{Tr}\left[ \gamma ^{\nu }\widetilde{S}_{c}^{ba^{\prime
}}(x)\gamma ^{\mu }S_{c}^{ab^{\prime }}(x)\right] \right\} .
\label{eq:QCDside1}
\end{eqnarray}

\begin{widetext}

\begin{figure}[h!]
\begin{center}
\includegraphics[totalheight=6cm,width=8cm]{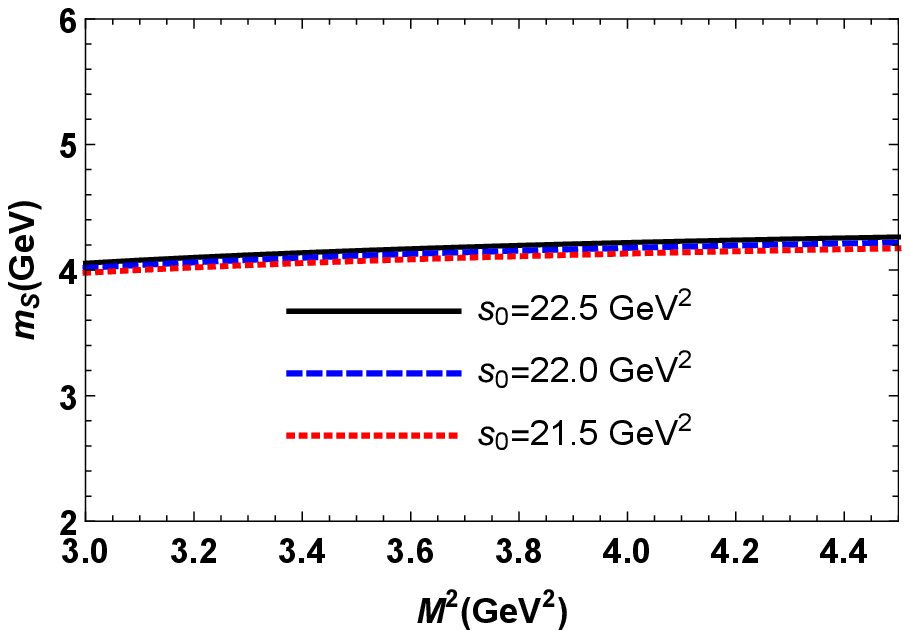}
\includegraphics[totalheight=6cm,width=8cm]{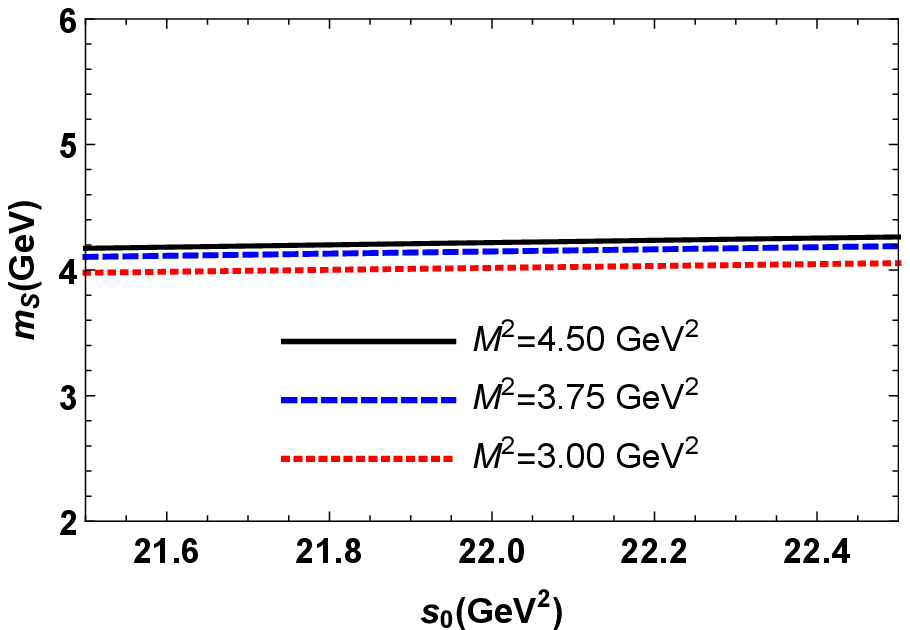}
\end{center}
\caption{Dependence of the $T_{\mathrm{S}}$ tetraquark's mass  $m_{\mathrm{S}}$  on the Borel parameter $M^{2}$ (left), and on the continuum threshold parameter $s_0$ (right).}
\label{fig:Mass2}
\end{figure}

\end{widetext}

Because the remaining operations with the correlation function $\Pi ^{%
\mathrm{OPE}}(p)$ are standard ones, we omit the further details and provide
only the final results for the mass $m_{\mathrm{S}}$ and coupling $f_{%
\mathrm{S}}$:%
\begin{eqnarray}
m_{\mathrm{S}} &=&(4128\pm 142)~\mathrm{MeV},  \notag \\
f_{\mathrm{S}} &=&(3.84\pm 0.34)\times 10^{-3}~\mathrm{GeV}^{4}.
\label{eq:ResultsScalar}
\end{eqnarray}%
Let us note that the values of $m_{\mathrm{S}}$ and $f_{\mathrm{S}}$ are
found utilizing for the parameters $M^{2}$ and $s_{0}$ working regions
\begin{equation}
M^{2}\in \lbrack 3,4.5]~\mathrm{GeV}^{2},\ s_{0}\in \lbrack 21.5,22.5]~%
\mathrm{GeV}^{2}.  \label{eq:Regions2}
\end{equation}%
In these regions the pole contribution varies within boundaries
\begin{equation}
0.69\geq \mathrm{PC}\geq 0.51,
\end{equation}%
and $R(3~\mathrm{GeV}^{2})<0.01$. The behavior of $m_{\mathrm{S}}$ as a
function of $M^{2}$ and $s_{0}$ is depicted in Fig.\ \ref{fig:Mass2}.

Similar analysis for the tetraquark $\widetilde{T}_{\mathrm{S}}$ gives
\begin{eqnarray}
&&\widetilde{\Pi }^{\mathrm{OPE}}(p)=i\int d^{4}xe^{ipx}\mathrm{Tr}\left[
\gamma _{5}\widetilde{S}_{s}^{b^{\prime }b}(-x)\right.  \notag \\
&&\left. \times \gamma _{5}S_{u}^{a^{\prime }a}(-x)\right] \left\{ \mathrm{Tr%
}\left[ \gamma _{5}\widetilde{S}_{c}^{aa^{\prime }}(x)\gamma
_{5}S_{c}^{bb^{\prime }}(x)\right] \right. \\
&&\left. +\mathrm{Tr}\left[ \gamma _{5}\widetilde{S}_{c}^{ba^{\prime
}}(x)\gamma _{5}S_{c}^{ab^{\prime }}(x)\right] \right\}.  \label{eq:QCDSide2}
\end{eqnarray}%
The spectroscopic parameters of $\widetilde{T}_{\mathrm{S}}$ are equal to
\begin{eqnarray}
\widetilde{m}_{\mathrm{S}} &=&(4035\pm 145)~\mathrm{MeV},  \notag \\
\widetilde{f}_{\mathrm{S}} &=&(7.7\pm 1.2)\times 10^{-3}~\mathrm{GeV}^{4}.
\label{eq:Results1a}
\end{eqnarray}%
To extract $\widetilde{m}_{\mathrm{S}}$ and $\widetilde{f}_{\mathrm{S}}$ ,
we used the windows Eq.\ (\ref{eq:Regions}) for the parameters $M^{2}$ and $%
s_{0}$, which satisfy all of the sum rule constraints in this case as well.


\section{Width of the tetraquark $T_{\mathrm{AV}}$}

\label{sec:Width1}
In the previous section, we have calculated the masses and couplings of the
tetraquarks $T_{\mathrm{AV}}$, $T_{\mathrm{S}}$ and $\widetilde{T}_{\mathrm{S%
}}$. This information forms a basis to determine the kinematically allowed
decay channels of these particles. In the case of the tetraquark $T_{\mathrm{%
AV}}$ the channels $T_{\mathrm{AV}}\rightarrow D^{0}D_{s}^{\ast +}$ and $T_{%
\mathrm{AV}}\rightarrow D^{\ast }(2007)^{0}D_{s}^{+}$ are its $S$-wave
modes, thresholds for which equal to $\approx 3975~\mathrm{MeV}$ and $3977~%
\mathrm{MeV}$, respectively. We calculate the full width of $T_{\mathrm{AV}}$
by taking into account these two channels.

It is worth noting, that we use central value of the mass $m$. But sum rule
computations depend on auxiliary parameters $M^{2}$ and $s_{0}$, which
restrict precision of the method by generating uncertainties in extracted
physical observables. Thus, the maximal predicted value for $m$ is $4138~%
\mathrm{MeV}$ which, however does not affect considerably our result for its
full widths. The reason is that even for such $m$ other strong decays of $T_{%
\mathrm{AV}}$, for instance, $P$-wave modes $T_{\mathrm{AV}}\rightarrow
D^{0}D_{s0}^{\ast }(2317)^{+}$ or $T_{\mathrm{AV}}\rightarrow D_{0}^{\ast
}(2400)^{0}D_{s}^{+}$ are kinematically forbidden processes because $4138~%
\mathrm{MeV}$ is below relevant two-meson thresholds.

At the $m$ less than the limit $3975~\mathrm{MeV}$ the axial-vector state $%
T_{\mathrm{AV}}$ cannot decay to $D^{0}D_{s}^{\ast +}$ and $D^{\ast
}(2007)^{0}D_{s}^{+}$. The process $T_{\mathrm{AV}}\rightarrow
D^{0}D^{0}K^{+}$, analog of $T_{cc}^{+}\rightarrow D^{0}D^{0}\pi ^{+}$,
could not also take place: A threshold for this decay $4224~\mathrm{MeV}$ is
significantly larger than $3975~\mathrm{MeV}$. Therefore, at $m<3975~\mathrm{%
MeV}$ the tetraquark $T_{\mathrm{AV}}$ becomes a strong-interaction stable
particle, which is excluded by the experimental data of the LHCb
Collaboration and theoretical analyses of diquark masses. In fact, the
axial-vector states $T_{cc}^{+}=cc\overline{u}\overline{d}$ and $T_{\mathrm{%
AV}}=cc\overline{u}\overline{s}$ contain the same heavy diquark $%
c^{T}C\gamma _{\mu }c$ and "good" light scalar antidiquarks $\overline{u}%
\gamma _{5}C\overline{d}^{T}$ and $\overline{u}\gamma _{5}C\overline{s}^{T}$
in $(\mathbf{3}_{f},\mathbf{3}_{c})$ flavor-color representations,
respectively. The mass splitting $\delta $ between "good" $J^{\mathrm{P}%
}=0^{+}$ diquarks $us$ and $ud$ was analyzed in Ref.\ \cite{Burden:1996nh}
and found equal to $145~\mathrm{MeV}$. This gap leads to the mass of $T_{%
\mathrm{AV}}$ approximately around of $4020~\mathrm{MeV}$. Our prediction
for $m$, as well as for the $T_{\mathrm{AV}}-T_{cc}^{+}$ mass splitting $%
\Delta =120~\mathrm{MeV}$ are close to these estimates. Thus, the
experimental-theoretical analyses rule out the small mass region for $T_{%
\mathrm{AV}}$, and confirm its unstability against strong decays to
conventional open-charmed mesons.


\subsection{Decay $T_{\mathrm{AV}}\rightarrow D^{0}D_{s}^{\ast +}$}


Here, we consider in a detailed form the process $T_{\mathrm{AV}}\rightarrow
D^{0}D_{s}^{\ast +}$. The partial width of this decay contains various
physical parameters of the initial and final particles, such as their masses
and decay constants. These parameters are known from other sources or have
been computed in the present article.

The partial width depends also on the strong interaction's coupling $g_{1}$
of the corresponding tetraquark and mesons at the vertex $T_{\mathrm{AV}%
}D^{0}D_{s}^{\ast +}$. It is convenient to determine $g_{1}$ by means of the
QCD three-point sum rule method. To this end, we analyze the correlation
function
\begin{eqnarray}
&&\Pi _{\mu \nu }(p,p^{\prime })=i^{2}\int d^{4}xd^{4}ye^{i(p^{\prime
}y-px)}\langle 0|\mathcal{T}\{J_{\nu }^{D_{s}^{\ast }}(y)  \notag \\
&&\times J^{D^{0}}(0)J_{\mu }^{\dagger }(x)\}|0\rangle ,  \label{eq:CF2}
\end{eqnarray}%
where $J_{\mu }(x)$,$\ J_{\nu }^{D_{s}^{\ast }}(y)$ and $J^{D^{0}}(0)$ are
the interpolating currents for the tetraquark $T_{\mathrm{AV}}$, the vector $%
D_{s}^{\ast +}$\ and pseudoscalar $D^{0}$ mesons, respectively. The
four-momenta of $T_{\mathrm{AV}}$ and $D_{s}^{\ast +}$ are denoted by $p$
and $p^{\prime }$, then momentum of the meson $D^{0}$ is equal to $%
q=p-p^{\prime }$.

The $J_{\mu }(x)$ is determined by Eq.\ (\ref{eq:CurrVector}), whereas for
currents of two final-state mesons, we use
\begin{eqnarray}
\ J_{\nu }^{D_{s}^{\ast }}(x) &=&\overline{s}_{j}(x)\gamma _{\nu }c_{j}(x),
\notag \\
J^{D^{0}}(x) &=&\overline{u}_{i}(x)i\gamma _{5}c_{i}(x),  \label{eq:Curr5}
\end{eqnarray}%
where $i$ and $j$ are color indices.

To continue our study in the framework of the sum rule method, we calculate
the correlation function $\Pi _{\mu \nu }(p,p^{\prime })$ by employing the
physical parameters of the particles involved into this process. For the
physical side of the sum rule, $\Pi _{\mu \nu }^{\mathrm{Phys}}(p,p^{\prime
})$, we obtain%
\begin{eqnarray}
&&\Pi _{\mu \nu }^{\mathrm{Phys}}(p,p^{\prime })=\frac{\langle 0|J_{\nu
}^{D^{\ast }}|D_{s}^{\ast +}(p^{\prime },\varepsilon ^{\prime })\rangle
\langle 0|J^{D^{0}}|D^{0}(q)\rangle }{(p^{\prime 2}-m_{D_{s}^{\ast
}}^{2})(q^{2}-m_{D}^{2})}  \notag \\
&&\times \frac{\langle D^{0}(q)D_{s}^{\ast +}(p^{\prime },\varepsilon
^{\prime })|T_{\mathrm{AV}}(p,\epsilon )\rangle \langle T_{\mathrm{AV}%
}(p,\epsilon )|J_{\mu }^{\dagger }|0\rangle }{(p^{2}-m^{2})}+\cdots ,  \notag
\\
&&  \label{eq:CF3}
\end{eqnarray}%
where $m_{D_{s}^{\ast }}$ and $m_{D}$ are the masses of the mesons $%
D_{s}^{\ast +}$ and $D^{0}$, respectively. To derive this expression, we
separate the contributions of the ground-state particles from the effects of
the higher resonances and continuum states in Eq.\ (\ref{eq:CF2}). Hence, in
Eq.\ (\ref{eq:CF3}), the ground-state term is written down explicitly,
whereas ellipses stand for the other contributions.

The function $\Pi _{\mu \nu }^{\mathrm{Phys}}(p,p^{\prime })$ can be
simplified by introducing the $D_{s}^{\ast +}$ and $D^{0}$ mesons' matrix
elements
\begin{eqnarray}
\langle 0|J_{\nu }^{D^{\ast }}|D_{s}^{\ast +}(p^{\prime },\varepsilon
^{\prime })\rangle &=&m_{D_{s}^{\ast }}f_{D_{s}^{\ast }}\varepsilon _{\nu
}^{\prime },\   \notag \\
\langle 0|J^{D^{0}}|D^{0}\rangle &=&\frac{m_{D}^{2}f_{D}}{m_{c}},
\label{eq:Mel2}
\end{eqnarray}%
with $f_{D_{s}^{\ast }}$ and $f_{D}$ being their decay constants. Here, $%
\varepsilon _{\nu }^{\prime }$ is the polarization vector of the meson $%
D_{s}^{\ast +}$.

The matrix element of the vertex $T_{\mathrm{AV}}D^{0}D_{s}^{\ast +}$ can be
modeled in the form
\begin{eqnarray}
&&\langle D^{0}(q)D_{s}^{\ast +}(p^{\prime },\varepsilon ^{\prime })|T_{%
\mathrm{AV}}(p,\epsilon )\rangle =g_{1}(q^{2})\left[ \left( p\cdot p^{\prime
}\right) \left( \epsilon \cdot \varepsilon ^{\prime \ast }\right) \right.
\notag \\
&&\left. -\left( p^{\prime }\cdot \epsilon \right) \left( p\cdot \varepsilon
^{\prime \ast }\right) \right] ,  \label{eq:Ver1}
\end{eqnarray}%
where we denote the strong coupling at $T_{\mathrm{AV}}D^{0}D_{s}^{\ast +}$
by $g_{1}(q^{2})$. Then, one can easily find that
\begin{eqnarray}
&&\Pi _{\mu \nu }^{\mathrm{Phys}}(p,p^{\prime })=g_{1}(q^{2})\frac{%
m_{D_{s}^{\ast }}f_{D_{s}^{\ast }}m_{D}^{2}f_{D}fm}{%
m_{c}(p^{2}-m^{2})(q^{2}-m_{D}^{2})}  \notag \\
&&\times \frac{1}{(p^{\prime 2}-m_{D_{s}^{\ast }}^{2})}\left( \frac{%
m^{2}+m_{D_{s}^{\ast }}^{2}-q^{2}}{2}g_{\mu \nu }-p_{\nu }p_{\mu }^{\prime
}\right) +\cdots .  \notag \\
&&  \label{eq:Phys2}
\end{eqnarray}%
The double Borel transformation of the correlation function over the
variables $-p^{2}$ and $-p^{\prime 2}$ is determined by the expression
\begin{eqnarray}
&&\mathcal{B}\Pi _{\mu \nu }^{\mathrm{Phys}}(p,p^{\prime })=g_{1}(q^{2})%
\frac{m_{D_{s}^{\ast }}f_{D_{s}^{\ast }}m_{D}^{2}f_{D}fm}{%
m_{c}(q^{2}-m_{D}^{2})}e^{-m^{2}/M_{1}^{2}}  \notag \\
&&\times e^{-m_{D_{s}^{\ast }}^{2}/M_{2}^{2}}\left( \frac{%
m^{2}+m_{D_{s}^{\ast }}^{2}-q^{2}}{2}g_{\mu \nu }-p_{\mu }p_{\nu }^{\prime
}\right) +\cdots .  \notag \\
&&
\end{eqnarray}%
The function $\mathcal{B}\Pi _{\mu \nu }^{\mathrm{Phys}}(p,p^{\prime })$ has
two Lorentz structures which are proportional to $g_{\mu \nu }$ and $p_{\nu
}p_{\mu }^{\prime }$. In our analysis, we work with the Borel transformation
of the invariant amplitude $\Pi ^{\mathrm{Phys}}(p^{2},p^{\prime 2},q^{2})$
which corresponds to the structure $\sim g_{\mu \nu }$.

To derive the QCD side of the three-point sum rule, we calculate $\Pi _{\mu
\nu }(p,p^{\prime })$ using the quark propagators, and obtain
\begin{eqnarray}
&&\Pi _{\mu \nu }^{\mathrm{OPE}}(p,p^{\prime })=i^{3}\int
d^{4}xd^{4}ye^{i(p^{\prime }y-px)}  \notag \\
&&\times \left\{ \mathrm{Tr}\left[ \gamma _{\nu }S_{c}^{ja}(y-x)\gamma _{\mu
}\widetilde{S}_{c}^{ib}(-x)\gamma _{5}\widetilde{S}_{u}^{ai}(x)\gamma
_{5}\right. \right.  \notag \\
&&\left. \times S_{s}^{bj}(x-y)\right] -\mathrm{Tr}\left[ \gamma _{\nu
}S_{c}^{jb}(y-x)\gamma _{\mu }\widetilde{S}_{c}^{ia}(-x)\right.  \notag \\
&&\left. \left. \times \gamma _{5}\widetilde{S}_{u}^{ai}(x)\gamma
_{5}S_{s}^{bj}(x-y)\right] \right\}.  \label{eq:CF4}
\end{eqnarray}

The correlation function $\Pi _{\mu \nu }^{\mathrm{OPE}}(p,p^{\prime })$ is
computed by taking into account the nonperturbative contributions up to
dimension $6$. It contains the same Lorentz structures as $\Pi _{\mu \nu }^{%
\mathrm{Phys}}(p,p^{\prime })$. Let us denote by $\Pi ^{\mathrm{OPE}%
}(p^{2},p^{\prime 2},q^{2})$ the invariant amplitude that corresponds to the
term proportional to $g_{\mu \nu }$ in $\Pi _{\mu \nu }^{\mathrm{OPE}%
}(p,p^{\prime })$. The double Borel transformation, $\mathcal{B}\Pi ^{%
\mathrm{OPE}}(p^{2},p^{\prime 2},q^{2})$, establishes the second component
of the sum rule. Having equated $\mathcal{B}\Pi ^{\mathrm{OPE}%
}(p^{2},p^{\prime 2},q^{2})$ and $\mathcal{B}\Pi ^{\mathrm{Phys}%
}(p^{2},p^{\prime 2},q^{2})$, and carried out the continuum subtraction, we
derive the sum rule for the strong coupling $g_{1}(q^{2})$.

The Borel transformed and subtracted amplitude $\Pi ^{\mathrm{OPE}%
}(p^{2},p^{\prime 2},q^{2})$ can be expressed by means of the spectral
density $\widetilde{\rho }(s,s^{\prime },q^{2})$ which is proportional to
the relevant imaginary part of the $\Pi _{\mu \nu }^{\mathrm{OPE}%
}(p,p^{\prime })$
\begin{eqnarray}
&&\Pi (\mathbf{M}^{2},\mathbf{s}_{0},q^{2})=\int_{\mathcal{M}%
^{2}}^{s_{0}}ds\int_{(m_{c}+m_{s})^{2}}^{s_{0}^{\prime }}ds^{\prime }\rho
(s,s^{\prime },q^{2})  \notag \\
&&\times e^{-s/M_{1}^{2}}e^{-s^{\prime }/M_{2}^{2}}.  \label{eq:SCoupl}
\end{eqnarray}%
The Borel and continuum threshold parameters are denoted in Eq.\ (\ref%
{eq:SCoupl}) by $\mathbf{M}^{2}=(M_{1}^{2},\ M_{2}^{2})$ and $\mathbf{s}%
_{0}=(s_{0},\ s_{0}^{\prime })$, respectively. Then, the sum rule for $%
g_{1}(q^{2})$ reads
\begin{eqnarray}
&&g_{1}(q^{2})=\frac{2m_{c}}{m_{D_{s}^{\ast }}f_{D_{s}^{\ast
}}m_{D}^{2}f_{D}fm}\frac{q^{2}-m_{D}^{2}}{m^{2}+m_{D_{s}^{\ast }}^{2}-q^{2}}
\notag \\
&&\times e^{m^{2}/M_{1}^{2}}e^{m_{D_{s}^{\ast }}^{2}/M_{2}^{2}}\Pi (\mathbf{M%
}^{2},\mathbf{s}_{0},q^{2}).  \label{eq:SRCoup}
\end{eqnarray}%
The coupling $g_{1}(q^{2})$ is also a function of the Borel and continuum
threshold parameters which for simplicity are omitted in Eq.\ (\ref%
{eq:SRCoup}). We also introduce a new variable $Q^{2}=-q^{2}$ and denote the
obtained function $g_{1}(Q^{2})$.

It is seen that Eq.\ (\ref{eq:SRCoup}) contains the mass and coupling of the
tetraquark $T_{\mathrm{AV}}$ as well as the masses and decay constants of
the mesons $D^{0}$ and $D_{s}^{\ast +}$. The relevant input parameters are
collected in Table\ \ref{tab:Param}, which contains also the masses and
decay constants of the mesons appearing at final stages of other processes.
The masses of all mesons and decay constants $f_{D}$ and $f_{D_{s}}$ are
borrowed from Ref.\ \cite{PDG:2022}. For decay constants of the mesons $%
D_{s}^{\ast +}$ and $D^{\ast }(2007)^{0}$, we use predictions obtained in
the context of the QCD lattice method \cite{Lubicz:2016bbi}.

Besides, for numerical analysis, we should determine working regions for
parameters $\mathbf{M}^{2}$ and $\mathbf{s}_{0}$. The restrictions imposed
on $\mathbf{M}^{2}$ and $\mathbf{s}_{0}$ are standard for sum rule
computations and have been discussed above. The windows for $M_{1}^{2}$ and $%
s_{0}$ correspond to the $T_{\mathrm{AV}}$ channel and coincide with regions
from Eq.\ (\ref{eq:Regions}). The second pair of parameters $(M_{2}^{2},\
s_{0}^{\prime })$ for the $D_{s}^{\ast +}$ channel are fixed within
boundaries:
\begin{equation}
M_{2}^{2}\in \lbrack 2.5,3.5]~\mathrm{GeV}^{2},\ s_{0}^{\prime }\in \lbrack
6,8]~\mathrm{GeV}^{2}.  \label{eq:Wind3}
\end{equation}%
The regions for the Borel and continuum subtraction parameters are chosen in
such a way that to minimize also dependence of $g_{1}(Q^{2})$ on them.

\begin{figure}[h]
\includegraphics[width=8.5cm]{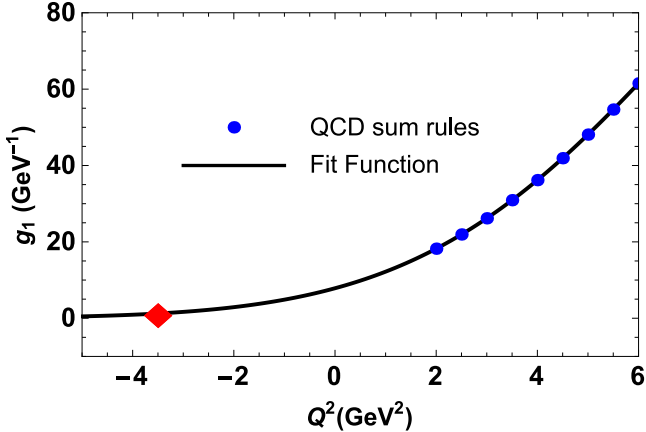}
\caption{The sum rule predictions and fit function for the strong coupling $%
g_{1}(Q^{2})$. The red diamond shows the point $Q^{2}=-m_{D}^{2}$. }
\label{fig:Fit}
\end{figure}

The width of the decay $T_{\mathrm{AV}}\rightarrow D^{0}D_{s}^{\ast +}$ has
to be calculated by means of the strong coupling at the $D^{0}$ meson's mass
shell $q^{2}=m_{D}^{2}$, which is not calculable by the sum rule method. To
avoid this obstacle, we use a fit function $\mathcal{F}_{i}(Q^{2})$ which at
the momenta $Q^{2}>0$ coincides with results of the sum rule analyses, but
can be extended to the $Q^{2}<0$ domain to find $g_{1}(-m_{D}^{2})$. For
these purposes, we employ the function $\mathcal{F}_{i}(Q^{2})$
\begin{equation}
\mathcal{F}_{i}(Q^{2})=\mathcal{F}_{i}^{0}\mathrm{\exp }\left[ c_{i}^{1}%
\frac{Q^{2}}{m^{2}}+c_{i}^{2}\left( \frac{Q^{2}}{m^{2}}\right) ^{2}\right] ,
\label{eq:FitF}
\end{equation}%
where $\mathcal{F}_{i}^{0}$, $c_{i}^{1}$ and $c_{i}^{2}$ are fitting
parameters. Numerical computations show that $\mathcal{F}_{1}^{0}=7.86~%
\mathrm{GeV}^{-1}$, $c_{1}^{1}=7.33$, and $c_{1}^{2}=-4.94$ lead to nice
agreement with the sum rule's data (see, Fig.\ \ref{fig:Fit}).

At the mass shell, $q^{2}=m_{D}^{2}$, this function predicts
\begin{equation}
g_{1}\equiv \mathcal{F}_{1}(-m_{D}^{2})=(1.26\pm 0.14)~\mathrm{GeV}^{-1}.
\label{eq:Coupl1}
\end{equation}%
The width of the process $T_{\mathrm{AV}}\rightarrow D^{0}D_{s}^{\ast +}$ is
determined by the following formula:%
\begin{equation}
\Gamma \left[ T_{\mathrm{AV}}\rightarrow D^{0}D_{s}^{\ast +}\right] =\frac{%
g_{1}^{2}m_{D_{s}^{\ast }}^{2}\lambda }{24\pi }\left( 3+\frac{2\lambda ^{2}}{%
m_{D_{s}^{\ast }}^{2}}\right) ,  \label{eq:PartDW}
\end{equation}%
where $\lambda =\lambda \left( m,m_{D_{s}^{\ast }},m_{D}\right) $ and
\begin{eqnarray}
\lambda \left( a,b,c\right) &=&\frac{1}{2a}\left[ a^{4}+b^{4}+c^{4}\right.
\notag \\
&&\left. -2\left( a^{2}b^{2}+a^{2}c^{2}+b^{2}c^{2}\right) \right] ^{1/2}.
\end{eqnarray}%
Employing the coupling $g_{1}$ from Eq.\ (\ref{eq:Coupl1}), it is easy to
find the partial width of the process $T_{\mathrm{AV}}\rightarrow
D^{0}D_{s}^{\ast +}$%
\begin{equation}
\Gamma \left[ T_{\mathrm{AV}}\rightarrow D^{0}D_{s}^{\ast +}\right] =(53\pm
12)~\mathrm{MeV}.  \label{eq:DW1Numeric}
\end{equation}

\subsection{Process $T_{\mathrm{AV}}\rightarrow D^{\ast }(2007)^{0}D_{s}^{+}$%
}


The second decay $T_{\mathrm{AV}}\rightarrow D^{\ast }(2007)^{0}D_{s}^{+}$
can be explored by the same manner. The correlation function which should be
considered in this case is
\begin{eqnarray}
\widetilde{\Pi }_{\mu \nu }(p,p^{\prime }) &=&i^{2}\int
d^{4}xd^{4}ye^{i(p^{\prime }y-px)}\langle 0|\mathcal{T}\{J^{D_{s}}(y)  \notag
\\
&&\times J_{\nu }^{D^{\ast 0}}(0)J_{\mu }^{\dagger }(x)\}|0\rangle ,
\label{eq:CF2a}
\end{eqnarray}%
where $\ J_{\nu }^{D^{\ast 0}}(0)$ and $J^{D_{s}}(y)$ are the interpolating
currents of the mesons $D^{\ast }(2007)^{0}$ and $D_{s}^{+}$, respectively.
These currents are determined by the formulas
\begin{eqnarray}
J_{\nu }^{D^{\ast 0}}(x) &=&\overline{u}_{i}(x)\gamma _{\nu }c_{i}(x),
\notag \\
J^{D_{s}}(x) &=&\overline{s}_{j}(x)i\gamma _{5}c_{j}(x).  \label{eq:Curr5a}
\end{eqnarray}%
To find the physical side of the sum rule, we use recipes described above
and get
\begin{eqnarray}
&&\widetilde{\Pi }_{\mu \nu }^{\mathrm{Phys}}(p,p^{\prime })=g_{2}(q^{2})%
\frac{m_{D^{\ast 0}}f_{D^{\ast 0}}m_{D_{s}}^{2}f_{D_{s}}fm}{%
(m_{c}+m_{s})(p^{2}-m^{2})(q^{2}-m_{D^{\ast 0}}^{2})}  \notag \\
&&\times \frac{1}{(p^{\prime 2}-m_{D_{s}}^{2})}\left( \frac{%
m^{2}-m_{D_{s}}^{2}+q^{2}}{2}g_{\mu \nu }-q_{\mu }p_{\nu }\right) +\cdots ,
\notag \\
&&  \label{eq:PhysSide4}
\end{eqnarray}%
where $m_{D^{\ast 0}}$, $m_{D_{s}}$ and $f_{D^{\ast 0}}$, $f_{D_{s}}$ are
the masses and decay constants of the mesons $D^{\ast }(2007)^{0}$ and $%
D_{s}^{+}$, respectively.

In deriving of Eq.\ (\ref{eq:PhysSide4}), we have used the following matrix
elements:%
\begin{eqnarray}
&&\langle 0|J_{\nu }^{D^{\ast 0}}|D^{\ast 0}(q,\varepsilon ^{\prime
})\rangle =m_{D^{\ast 0}}f_{D^{\ast 0}}\varepsilon _{\nu }^{\prime },\
\notag \\
&&\langle 0|J^{D_{s}}|D_{s}^{+}\rangle =\frac{m_{D_{s}}^{2}f_{D_{s}}}{%
m_{c}+m_{s}},  \notag \\
&&\langle D^{\ast 0}(q,\varepsilon ^{\prime })D_{s}^{+}(p^{\prime })|T_{%
\mathrm{AV}}(p,\epsilon )\rangle =g_{2}(q^{2})\left[ \left( p\cdot q\right)
\left( \epsilon \cdot \varepsilon ^{\prime \ast }\right) \right.  \notag \\
&&\left. -\left( q\cdot \epsilon \right) \left( p\cdot \varepsilon ^{\prime
\ast }\right) \right].  \label{eq:Mel2a}
\end{eqnarray}%
Here $g_{2}$ is the strong coupling, which corresponds to the vertex $T_{%
\mathrm{AV}}D^{\ast }(2007)^{0}D_{s}^{+}$ and is defined at the mass shell
of the $D^{\ast }(2007)^{0}$ meson,
\begin{equation}
g_{2}=\mathcal{F}_{2}(-m_{D^{\ast 0}}^{2}).
\end{equation}%
The fit function $\mathcal{F}_{2}(Q^{2})$ is given by Eq.\ (\ref{eq:FitF})
with the parameters: $\mathcal{F}_{2}^{0}=5.36~\mathrm{GeV}^{-1}$, $%
c_{2}^{1}=3.68$, and $c_{2}^{2}=-15.73$. The partial width of this decay can
be calculated by means of the formula in Eq.\ (\ref{eq:PartDW}) with evident
replacements $g_{1}\rightarrow g_{2}$, $m_{D_{s}^{\ast }}^{2}\rightarrow
m_{D^{\ast 0}}^{2}$ and $\lambda \left( m,m_{D_{s}^{\ast }},m_{D}\right)
\rightarrow \lambda \left( m,m_{D^{\ast 0}},m_{D_{s}}\right)$.

 In the  sum rule computations, we use  Eq.\ (\ref{eq:Regions2}) and $ M^{2}_2 \in [2.5,3.5]~\mathrm{GeV}^{2} $, $ s'_0\in [5,7]~\mathrm{GeV}^{2} $. As a result, we find
\begin{equation}
g_{2}\equiv \mathcal{F}_{2}(-m_{D^{\ast 0}}^{2})=(7.8\pm 0.9)\cdot 10^{-1}\
\mathrm{GeV}^{-1}.
\end{equation}%
Then, the width of the process $T_{\mathrm{AV}}\rightarrow D^{\ast
}(2007)^{0}D_{s}^{+}$ is equal to
\begin{equation}
\Gamma \left[ T_{\mathrm{AV}}\rightarrow D^{\ast }(2007)^{0}D_{s}^{+}\right]
=(19\pm 5)~\mathrm{MeV}.  \label{eq:DW2Numeric}
\end{equation}%
For the full width of the exotic axial-vector meson $T_{\mathrm{AV}}$, we
get
\begin{equation}
\Gamma _{\mathrm{AV}}=(72\pm 13)~\mathrm{MeV}.  \label{eq:FullW}
\end{equation}

\begin{table}[tbp]
\begin{tabular}{|c|c|}
\hline\hline
Quantity & Value (in $\mathrm{MeV}$ units) \\ \hline
$m_{D_{s}^{\ast}}$ & $2112.2\pm 0.4$ \\
$m_{D}$ & $1864.84\pm 0.05$ \\
$m_{D_{s}}$ & $1969.0 \pm 1.4$ \\
$m_{D^{\ast 0}}$ & $2006.85\pm 0.05$ \\
$f_{D_{s}^{\ast}}$ & $268.8 \pm 6.5$ \\
$f_{D}$ & $212.0 \pm 0.7$ \\
$f_{D_{s}}$ & $249.9 \pm 0.5$ \\
$f_{D^{\ast 0}}$ & $223.5 \pm 8.4$ \\ \hline\hline
\end{tabular}%
\caption{Masses and decay constants of the mesons $D_{s}^{\ast +}$, $D^{0}$,
$D_{s}^{+}$, and $D^{\ast }(2007)^{0}$, which have been used in numerical
computations.}
\label{tab:Param}
\end{table}
This result is rather stable prediction for $\Gamma _{\mathrm{AV}}$, because
in the region $3995~\mathrm{MeV}\leq m\leq 4138~\mathrm{MeV}$ there are not
other allowed decays of $T_{\mathrm{AV}}$, and its full width may undergone
only small variations due to changes of $m$ and $\lambda $ in Eq.\ (\ref%
{eq:PartDW}).


\section{Full widths of $T_{\mathrm{S}}$ and $\widetilde{T}_{\mathrm{S}}$}

\label{sec:Width2}
In the case of the scalar tetraquark $T_{\mathrm{S}}$, the $S$-wave decay
channels which contribute to its full width are the processes $T_{\mathrm{S}%
}\rightarrow D^{0}D_{s}^{+}$ and $T_{\mathrm{S}}\rightarrow D^{\ast
}(2007)^{0}D_{s}^{\ast +}$. The two-meson thresholds for these decays $3833~%
\mathrm{MeV}$ and $4119~\mathrm{MeV}$ make them dominant channels of the
tetraquark $T_{\mathrm{S}}$. There are other channels via of which the
scalar four-quark state $cc\overline{u}\overline{s}$ may transform to
conventional charmed mesons. For example, decays to meson pairs $D_{0}^{\ast
}(2400)^{0}D_{s0}^{\ast }(2317)^{+}$ and $D^{\ast }(2007)^{0}D_{s0}^{\ast
}(2317)^{+}$ are possible $S$- and $P$-wave modes of $cc\overline{u}%
\overline{s}$, respectively. But thresholds for production of these and
other pairs exceed considerably the maximal value $m_{\mathrm{S}}=4270~%
\mathrm{MeV}$ predicted for $T_{\mathrm{S}}$. In the case of $\widetilde{T}_{%
\mathrm{S}}$, we explore the decay $\widetilde{T}_{\mathrm{S}}\rightarrow
D^{0}D_{s}^{+}$ allowed by its mass $\widetilde{m}_{\mathrm{S}}=4035~\mathrm{%
MeV}$.


\subsection{Decay modes $T_{\mathrm{S}}\rightarrow D^{0}D_{s}^{+}$ and $T_{%
\mathrm{S}}\rightarrow D^{\ast }(2007)^{0}D_{s}^{\ast +}$}


The treatment of the decays $T_{\mathrm{S}}\rightarrow D^{0}D_{s}^{+}$ and $%
T_{\mathrm{S}}\rightarrow D^{\ast }(2007)^{0}D_{s}^{\ast +}$ in the context
of the three-point sum rule approach does differ from our analysis made in
the previous section. Here, one should find the strong couplings $G_{1}$ and
$G_{2}$ at the relevant vertices. For the coupling $G_{1}$ the correlation
function of interest is
\begin{eqnarray}
&&\Pi (p,p^{\prime })=i^{2}\int d^{4}xd^{4}ye^{i(p^{\prime }y-px)}\langle 0|%
\mathcal{T}\{J^{D_{s}}(y)  \notag \\
&&\times J^{D^{0}}(0)J^{\dagger }(x)\}|0\rangle ,  \label{eq:CorrF6}
\end{eqnarray}%
where all interpolating currents have been defined above. Thus, the current $%
J(x)$ of the scalar tetraquark $T_{\mathrm{S}}$ has been introduced in Eq.\ (%
\ref{eq:Curr1}), whereas $J^{D^{0}}(x)$ and $J^{D_{s}}(x)$ have been
determined by Eqs.\ (\ref{eq:Curr5}) and (\ref{eq:Curr5a}), respectively.

The physical side of the sum rule has the form
\begin{eqnarray}
&&\Pi ^{\mathrm{Phys}}(p,p^{\prime })=\frac{\langle
0|J^{D_{s}}|D_{s}^{+}(p^{\prime })\rangle \langle
0|J^{D^{0}}|D^{0}(q)\rangle }{(p^{\prime 2}-m_{D_{s}}^{2})(q^{2}-m_{D}^{2})}
\notag \\
&&\times \frac{\langle D^{0}(q)D_{s}^{+}(p^{\prime })|T_{\mathrm{S}%
}(p)\rangle \langle T_{\mathrm{S}}(p)|J^{\dagger }|0\rangle }{(p^{2}-m_{%
\mathrm{S}}^{2})}+\cdots .  \label{eq:CF3a}
\end{eqnarray}%
The matrix elements of the mesons $D^{0}$ and $D_{s}^{+}$ have been defined
by Eqs.\ (\ref{eq:Mel2}) and (\ref{eq:Mel2a}), respectively. We model the
matrix element $\langle D^{0}(q)D_{s}^{+}(p^{\prime })|T_{\mathrm{S}%
}(p)\rangle $ as
\begin{equation}
\langle D^{0}(q)D_{s}^{+}(p^{\prime })|T_{\mathrm{S}}(p)\rangle
=G_{1}(q^{2})p\cdot p^{\prime }.  \label{eq:Ver2}
\end{equation}%
These matrix elements allow us to rewrite $\Pi ^{\mathrm{Phys}}(p,p^{\prime
})$ in a simplified form,%
\begin{eqnarray}
&&\Pi ^{\mathrm{Phys}}(p,p^{\prime })=G_{1}(q^{2})\frac{%
m_{D}^{2}f_{D}m_{D_{s}}^{2}f_{D_{s}}f_{\mathrm{S}}m_{\mathrm{S}}}{%
m_{c}(m_{c}+m_{s})(p^{2}-m_{\mathrm{S}}^{2})}  \notag \\
&&\times \frac{m_{\mathrm{S}}^{2}+m_{D_{s}}^{2}-q^{2}}{%
2(q^{2}-m_{D}^{2})(p^{\prime 2}-m_{D_{s}}^{2})}+\cdots .  \label{eq:Phys3}
\end{eqnarray}%
As is seen, the correlation function $\Pi ^{\mathrm{Phys}}(p,p^{\prime })$
has a simple Lorentz structure proportional to $\mathrm{I}$, therefore we
employ the whole expression in Eq.\ (\ref{eq:Phys3}) as an invariant
amplitude to derive the sum rule for $G_{1}(q^{2})$.

After some calculations, we find also the QCD side of the required sum rule
\begin{eqnarray}
&&\Pi ^{\mathrm{OPE}}(p,p^{\prime })=\int d^{4}xd^{4}ye^{i(p^{\prime }y-px)}
\notag \\
&&\times \left\{ \mathrm{Tr}\left[ \gamma _{5}S_{c}^{ja}(y-x)\gamma _{\mu }%
\widetilde{S}_{c}^{ib}(-x)\gamma _{5}\widetilde{S}_{u}^{ai}(x)\gamma ^{\mu
}\right. \right.  \notag \\
&&\left. \times S_{s}^{bj}(x-y)\right] -\mathrm{Tr}\left[ \gamma
_{5}S_{c}^{jb}(y-x)\gamma _{\mu }\widetilde{S}_{c}^{ia}(-x)\right.  \notag \\
&&\left. \left. \times \gamma _{5}\widetilde{S}_{u}^{ai}(x)\gamma ^{\mu
}S_{s}^{bj}(x-y)\right] \right\} .  \label{eq:CF5a}
\end{eqnarray}%
Then the sum rule for the coupling $G_{1}(q^{2})$ reads%
\begin{eqnarray}
&&G_{1}(q^{2})=\frac{2m_{c}(m_{c}+m_{s})}{%
m_{D}^{2}f_{D}m_{D_{s}}^{2}f_{D_{s}}f_{\mathrm{S}}m_{\mathrm{S}}}\frac{%
q^{2}-m_{D}^{2}}{m_{\mathrm{S}}^{2}+m_{D_{s}}^{2}-q^{2}}  \notag \\
&&\times e^{m_{\mathrm{S}}^{2}/M_{1}^{2}}e^{m_{D_{s}}^{2}/M_{2}^{2}}\Pi _{%
\mathrm{scal}}(\mathbf{M}^{2},\mathbf{s}_{0},q^{2}),  \label{eq:SCG}
\end{eqnarray}%
where $\Pi _{\mathrm{scal}}(\mathbf{M}^{2},\mathbf{s}_{0},q^{2})$ is the
correlation function $\Pi ^{\mathrm{OPE}}(p,p^{\prime })$ after the Borel
transformation and subtraction operations. It is calculated by taking into
account the nonperturbative terms up to dimension $6$, as two similar
functions in the previous section.

Computations carried out in accordance with a scheme described above, give
the following predictions%
\begin{equation}
G_{1}=\mathcal{F}_{3}(-m_{D}^{2})=(1.01\pm 0.21)\ \mathrm{GeV}^{-1},
\end{equation}%
where parameters of the function $\mathcal{F}_{3}(Q^{2})$ are: $\mathcal{F}%
_{3}^{0}=4.10~\mathrm{GeV}^{-1}$, $c_{3}^{1}=4.86$, and $c_{3}^{2}=-9.90$. It is worth nothing that the fit function is given by  Eq.\ (\ref{eq:FitF}) with substitution $ m \rightarrow m_S $. In the  sum rule computations for the $ T_S $ channel, we employ the parameters from  Eq.\ (\ref{eq:Regions2}).

The partial width of this decay is determined by the expression
\begin{equation}
\Gamma \left[ T_{\mathrm{S}}\rightarrow D^{0}D_{s}^{+}\right] =\frac{%
G_{1}^{2}m_{D_{s}}^{2}\lambda }{8\pi }\left( 1+\frac{\lambda ^{2}}{%
m_{D_{s}}^{2}}\right) ,  \label{eq:SDWform}
\end{equation}%
with $\lambda $ being equal to $\lambda \left( m_{\mathrm{S}%
},m_{D_{s}},m_{D}\right) $. Numerical computations yield%
\begin{equation}
\Gamma \left[ T_{\mathrm{S}}\rightarrow D^{0}D_{s}^{+}\right] =(138\pm 41)~%
\mathrm{MeV}.  \label{eq:PWS1}
\end{equation}

To determine the coupling $G_{2}$ and study the decay $T_{\mathrm{S}%
}\rightarrow D^{\ast }(2007)^{0}D_{s}^{\ast +}$, we analyze the correlation
function
\begin{eqnarray}
&&\widehat{\Pi }_{\mu \nu }(p,p^{\prime })=i^{2}\int
d^{4}xd^{4}ye^{i(p^{\prime }y-px)}\langle 0|\mathcal{T}\{J_{\mu
}^{D_{s}^{\ast }}(y)  \notag \\
&&\times J_{\nu }^{D^{\ast 0}}(0)J^{\dagger }(x)\}|0\rangle ,
\end{eqnarray}%
with $J_{\mu }^{D_{s}^{\ast }}(y)$ and $J_{\nu }^{D^{\ast 0}}(0)$ being the
interpolating currents of the vector mesons $D_{s}^{\ast +}$ and $D^{\ast
}(2007)^{0}$ given by Eqs.\ (\ref{eq:Curr5}) and\ (\ref{eq:Curr5a}),
respectively.

The function $\widehat{\Pi }_{\mu \nu }(p,p^{\prime })$ in terms of the
physical parameters of the tetraquark $T_{\mathrm{S}}$ and mesons $%
D_{s}^{\ast +}$ and $D^{\ast }(2007)^{0}$ is equal to expression%
\begin{eqnarray}
&&\widehat{\Pi }_{\mu \nu }^{\mathrm{Phys}}(p,p^{\prime })=G_{2}(q^{2})\frac{%
m_{D_{s}^{\ast }}f_{D_{s}^{\ast }}m_{D^{\ast 0}}f_{D^{\ast 0}}f_{\mathrm{S}%
}m_{\mathrm{S}}}{(p^{2}-m_{\mathrm{S}}^{2})(p^{\prime 2}-m_{D_{s}^{\ast
}}^{2})}  \notag \\
&&\times \frac{1}{(q^{2}-m_{D^{\ast 0}}^{2})}\left( \frac{m_{\mathrm{S}%
}^{2}-m_{D_{s}^{\ast }}^{2}-q^{2}}{2}g_{\mu \nu }-q_{\mu }p_{\nu }^{\prime
}\right) +\cdots ,\notag \\
\end{eqnarray}%
where the vertex $\langle D^{\ast }(2007)^{0}(q,\varepsilon )D_{s}^{\ast
+}(p^{\prime },\varepsilon ^{\prime })|T_{\mathrm{S}}(p)\rangle $ is modeled
in the form%
\begin{eqnarray}
&&\langle D^{\ast }(2007)^{0}(q,\varepsilon )D_{s}^{\ast +}(p^{\prime
},\varepsilon ^{\prime })|T_{\mathrm{S}}(p)\rangle =G_{2}(q^{2})  \notag \\
&&\times \left[ \left( p^{\prime }\cdot q\right) \left( \varepsilon ^{\ast
}\cdot \varepsilon ^{\prime \ast }\right) -\left( q\cdot \varepsilon
^{\prime \ast }\right) \left( p^{\prime }\cdot \varepsilon ^{\ast }\right) %
\right] .
\end{eqnarray}%
In terms of the quark-gluon degrees of freedom, $\widehat{\Pi }_{\mu \nu }^{%
\mathrm{OPE}}(p,p^{\prime })$ is given by the formula
\begin{eqnarray}
&&\widehat{\Pi }_{\mu \nu }^{\mathrm{OPE}}(p,p^{\prime })=\int
d^{4}xd^{4}ye^{i(p^{\prime }y-px)}\left\{ \mathrm{Tr}\left[ \gamma _{\mu
}S_{c}^{ia}(y-x)\right. \right.  \notag \\
&&\left. \times \gamma _{\alpha }\widetilde{S}_{c}^{jb}(-x)\gamma _{\nu }%
\widetilde{S}_{s}^{bj}(x)\gamma ^{\alpha }S_{u}^{ai}(x-y)\right]  \notag \\
&&\left. -\mathrm{Tr}\left[ \gamma _{\mu }S_{c}^{ib}(y-x)\gamma _{\alpha }%
\widetilde{S}_{c}^{ja}(-x)\gamma _{\nu }\widetilde{S}_{s}^{bj}(x)\gamma
^{\alpha }S_{u}^{ai}(x-y)\right] \right\}.  \notag \\
&&  \label{eq:CF5b}
\end{eqnarray}

Omitting details of rather standard computations, let us write down the
final results:%
\begin{equation}
G_{2}=\mathcal{F}_{4}(-m_{D^{\ast 0}}^{2})=(2.11\pm 0.43)\ \mathrm{GeV}^{-1},
\end{equation}%
where parameters of the function $\mathcal{F}_{4}(Q^{2})$ are: $\mathcal{F}%
_{4}^{0}=3.17~\mathrm{GeV}^{-1}$, $c_{4}^{1}=1.14$, and $c_{4}^{2}=-2.51$.

The partial width of this decay is equal to
\begin{equation}
\Gamma \left[ T_{\mathrm{S}}\rightarrow D^{\ast }(2007)^{0}D_{s}^{\ast +}%
\right] =(75\pm 22)~\mathrm{MeV}.  \label{eq:PWS2}
\end{equation}%
Using Eqs.\ (\ref{eq:PWS1}) and\ (\ref{eq:PWS2}), we find
\begin{equation}
\Gamma _{\mathrm{S}}=(213\pm 47)~\mathrm{MeV}.
\end{equation}


\subsection{ Decay $\widetilde{T}_{\mathrm{S}}\rightarrow D^{0}D_{s}^{+}$}


For the decay $\widetilde{T}_{\mathrm{S}}\rightarrow D^{0}D_{s}^{+}$ one
should employ the following correlation function
\begin{eqnarray}
&&\widetilde{\Pi }^{\mathrm{OPE}}(p,p^{\prime })=-\int
d^{4}xd^{4}ye^{i(p^{\prime }y-px)}  \notag \\
&&\times \left\{ \mathrm{Tr}\left[ \gamma _{5}S_{c}^{ja}(y-x)\gamma _{5}%
\widetilde{S}_{c}^{ib}(-x)\gamma _{5}\widetilde{S}_{u}^{ai}(x)\gamma
_{5}\right. \right.  \notag \\
&&\left. \times S_{s}^{bj}(x-y)\right] +\mathrm{Tr}\left[ \gamma
_{5}S_{c}^{jb}(y-x)\gamma _{5}\widetilde{S}_{c}^{ia}(-x)\right.  \notag \\
&&\left. \left. \times \gamma _{5}\widetilde{S}_{u}^{ai}(x)\gamma
_{5}S_{s}^{bj}(x-y)\right] \right\}.  \label{eq:CF6}
\end{eqnarray}%
Then $\widetilde{G}(q^{2})$ can be extracted from the sum rule%
\begin{eqnarray}
&&\widetilde{G}(q^{2}) =\frac{2m_{c}(m_{c}+m_{s})}{%
m_{D}^{2}f_{D}m_{D_{s}}^{2}f_{D_{s}}\widetilde{f}_{\mathrm{S}}\widetilde{m}_{%
\mathrm{S}}}\frac{q^{2}-m_{D}^{2}}{\widetilde{m}_{\mathrm{S}%
}^{2}+m_{D_{s}}^{2}-q^{2}}  \notag \\
&&\times e^{\widetilde{m}_{\mathrm{S}%
}^{2}/M_{1}^{2}}e^{m_{D_{s}}^{2}/M_{2}^{2}}\widetilde{\Pi }_{\mathrm{scal}}(%
\mathbf{M}^{2},\mathbf{s}_{0},q^{2}),  \label{eq:CouplG}
\end{eqnarray}%
where $\widetilde{\Pi }_{\mathrm{scal}}(\mathbf{M}^{2},\mathbf{s}_{0},q^{2})$
is the Borel transformed and subtracted correlation function $\widetilde{\Pi
}^{\mathrm{OPE}}(p,p^{\prime })$.

Computations give the following predictions:%
\begin{equation}
\widetilde{G}=\mathcal{F}_{5}(-m_{D}^{2})=(1.07\pm 0.21)\ \mathrm{GeV}^{-1},
\end{equation}%
where parameters of the function $\mathcal{F}_{5}(Q^{2})$, obtained from Eq.\ (\ref{eq:FitF}) after the replacement $ m \rightarrow \tilde{m}_S$, are: $\mathcal{F}%
_{5}^{0}=4.06~\mathrm{GeV}^{-1}$, $c_{5}^{1}=4.56$, and $c_{5}^{2}=-7.82$.

The partial width of this decay is determined by the expression Eq.\ (\ref%
{eq:SDWform}) after substitutions $G_{1}\rightarrow \widetilde{G}$, $m_{%
\mathrm{S}}\rightarrow $ $\widetilde{m}_{\mathrm{S}}$ and $\lambda
\rightarrow $ $\widetilde{\lambda }=$ $\lambda (\widetilde{m}_{\mathrm{S}%
},m_{D_{s}},m_{D})$.

Numerical computations yield%
\begin{equation}
\widetilde{\Gamma }_{\mathrm{S}}=(123\pm 32)~\mathrm{MeV},  \label{eq:FullW2}
\end{equation}%
which characterizes $\widetilde{T}_{\mathrm{S}}$ as a wide resonance.


\section{Discussion and concluding notes}

\label{sec:Conclusion}

The main motivation for present investigation is the LHCb discovery of the
very narrow doubly charmed axial-vector state $T_{cc}^{+}$. It is special in
two respects: First, $T_{cc}^{+}$ is the only doubly charmed resonance
observed experimentally. Secondly, it is narrowest candidate to a four-quark
meson. These circumstances made $T_{cc}^{+}$ an object of intensive
theoretical studies: Its structure and parameters were investigated in
numerous publications using various methods and models.

This discovery generated interest to possible counterparts of $T_{cc}^{+}$,
which may be four-quark systems with the same content but different
spin-parities. The doubly charmed tetraquarks containing strange $s$%
-quark(s) are another class of particles closely related to $T_{cc}^{+}$. In
present article, we have concentrated namely on these particles and explored
the axial-vector and scalar doubly charmed strange tetraquarks $T_{\mathrm{AV%
}}$, $T_{\mathrm{S}}$, and $\widetilde{T}_{\mathrm{S}}$ by calculating their
masses and widths.

It is worth to compare predictions obtained for the mass $m=(3995\pm 143)~%
\mathrm{MeV}$ and width $\Gamma _{\mathrm{AV}}=(72\pm 13)~\mathrm{MeV}$ of $%
T_{\mathrm{AV}}$ with parameters of $T_{cc}^{+}$ measured by the LHCb
Collaboration. The mass gap between these two states, in accordance with our
findings, amounts to $120~\mathrm{MeV}$, which can be considered as a
reasonable estimate for particles with a $s$-quark difference in their
contents. The two-meson states $D^{0}D_{s}^{\ast +}$ and $D^{\ast
}(2007)^{0}D_{s}^{+}$ play for the tetraquark $T_{\mathrm{AV}}$ the same
role as $D^{0}D^{\ast }{}^{+}$ for the resonance $T_{cc}^{+}$. The mass of $%
T_{cc}^{+}$ is very close but less than $D^{0}D^{\ast }{}^{+}$ threshold,
whereas $T_{\mathrm{AV}}$ lies $\approx 20~\mathrm{MeV}$ above corresponding
thresholds, which makes its decays to mesons $D^{0}D_{s}^{\ast +}$ and $%
D^{\ast }(2007)^{0}D_{s}^{+}$ kinematically allowed processes. These $S$%
-wave channels form width of the tetraquark $T_{\mathrm{AV}}$, and our
prediction for $\Gamma _{\mathrm{AV}}$ means that it is relatively wide
resonance. In this aspect, $T_{\mathrm{AV}}$ differs from the very narrow
state $T_{cc}^{+}$, because $T_{\mathrm{AV}}$ and $T_{cc}^{+}$ decay to
ordinary mesons through different mechanisms. Thus, the decay mode $%
T_{cc}^{+}\rightarrow D^{0}D^{0}\pi ^{+}$ in which $T_{cc}^{+}$ was
discovered, may run due to the transformation $T_{cc}^{+}\rightarrow
D^{0}D^{\ast +}$. But this process is forbidden, therefore the final state $%
D^{0}D^{0}\pi ^{+}$ appears through production of intermediate scalar
tetraquarks, partial widths of which are very small.

Another interesting question to be addressed here is the mass splitting
between the axial-vector $T_{\mathrm{AV}}$ and scalar $T_{\mathrm{S}}$
tetraquarks. The $T_{\mathrm{AV}}$ contains $J^{\mathrm{P}}=0^{+}$ "good" $(%
\mathbf{3}_{f},\mathbf{3}_{c})$ antidiquark $\overline{u}\overline{s}$,
whereas $T_{\mathrm{S}}$ is made of $J^{\mathrm{P}}=1^{+}$ "bad" $(\overline{%
\mathbf{6}}_{f},\mathbf{3}_{c})$ one. For the nonstrange diquark the
"bad"-"good" mass difference $\delta (1^{+}-0^{+})_{ud}$ was found from the
QCD lattice calculations equal to $198~\mathrm{MeV}$ \cite{Francis:2021vrr},
which transforms to $\delta (1^{+}-0^{+})_{us}=135~\mathrm{MeV}$ in the case
of $\overline{u}\overline{s}$ diquark \cite{Karliner:2021wju}. Then, in
accordance with this scheme, $m_{\mathrm{S}}$ should be equal approximately
to $4130~\mathrm{MeV}$. As is seen, our sum rule prediction for $m_{\mathrm{S%
}}=(4128\pm 142)~\mathrm{MeV}$ agrees well with this estimate. The
tetraquark $\widetilde{T}_{\mathrm{S}}$ built of color-sextet scalar
diquarks resides between $T_{\mathrm{AV}}$ and $T_{\mathrm{S}}$ states. In a
situation when an existence of a scalar state $cc\overline{u}\overline{s}$
with "good" components is forbidden, the axial-vector tetraquark $T_{\mathrm{%
AV}}$ becomes the lightest particle in the spectrum.

The four-quark structure $cc\overline{u}\overline{s}$ with $J^{\mathrm{P}%
}=1^{+}$ was studied in other publications in the framework of alternative
approaches. In the diquark-antidiquark and molecule models its mass was
estimated as $4106~\mathrm{MeV}$ and $3974.8_{-0.5}^{+0.4}~\mathrm{MeV}$ in
Refs.\ \cite{Karliner:2021wju} and \cite{Ren:2021dsi}, respectively. As it
has been emphasized above, the prediction around of $3975~\mathrm{MeV}$
contradicts to experimental-theoretical constraints on the mass of $T_{%
\mathrm{AV}}$. Our result is less than the prediction made in Ref.\ \cite%
{Karliner:2021wju} though $m$ in upper limit overlaps with it.

One sees, that there are interesting open problems in physics of doubly
charmed four-quark mesons, which require additional theoretical and
experimental studies. The information gained in this article on the masses
and widths of the strange doubly-charmed tetraquarks gives new perspectives
on these states and can be used in future experimental investigations of
multiquark hadrons.

\textbf{Data Availability Statement:} No Data associated in the manuscript.

\end{document}